\documentclass{article} 
\usepackage{iclr2021_conference,times}

\usepackage{amsmath,amsfonts,bm}

\def\eqref#1{equation~\ref{#1}}

\def\1{\bm{1}}

\DeclareMathAlphabet{\mathsfit}{\encodingdefault}{\sfdefault}{m}{sl}
\SetMathAlphabet{\mathsfit}{bold}{\encodingdefault}{\sfdefault}{bx}{n}

\usepackage{hyperref}
\usepackage{graphicx}
\usepackage{url}

\title{Mobile Apps Prioritizing Privacy, Efficiency and Equity: A Decentralized Approach to COVID-19 Vaccination Coordination}

\author{\\\textbf{Joseph Bae}\textsuperscript{1, 3}, \textbf{Rohan Sukumaran}\textsuperscript{1}, \textbf{Sheshank Shankar}\textsuperscript{1},  \textbf{Anshuman Sharma}\textsuperscript{1}, \textbf{Ishaan Singh}\textsuperscript{1},\\ \textbf{Haris Nazir}\textsuperscript{1}, \textbf{Colin Kang}\textsuperscript{1}, \textbf{Saurish Srivastava}\textsuperscript{1},  \textbf{Parth Patwa}\textsuperscript{1}, \textbf{Priyanshi} \\ \textbf{Katiyar}\textsuperscript{1},
\textbf{Vitor Pamplona}\textsuperscript{1}  \\\\
\textsuperscript{1}PathCheck Foundation, 02139 Cambridge, USA.\\
\textsuperscript{2}MIT Media Lab, 02139 Cambridge, USA.\\
\textsuperscript{3}Renaissance School of Medicine, Stony Brook University, 11794 Stony Brook, USA.\\
\texttt{joseph.bae@stonybrookmedicine.edu} }

\iclrfinalcopy 
\begin{document}

\maketitle

\begin{abstract}
    In this early draft, we describe a decentralized, app-based approach to COVID-19 vaccine distribution that facilitates zero knowledge verification, dynamic vaccine scheduling, continuous symptoms reporting, access to aggregate analytics based on population trends and more. To ensure equity, our solution is developed to work with limited internet access as well. In addition, we describe the six critical functions that we believe last mile vaccination management platforms must perform, examine existing vaccine management systems, and present a model for privacy-focused, individual-centric solutions.
\end{abstract}

\section{Introduction}
The efficient and rapid distribution of COVID-19 vaccine doses demands that every step in the process is carefully planned to address logistical, communication, privacy and ethical challenges \cite{bae2020challenges}.

In this paper, we describe six critical functions that we believe last mile vaccine management platforms must perform. We examine existing vaccine distribution/management systems and present a model for privacy-focused, individual-centric solutions. 

Here, we propose a  decentralized, mobile app approach to  address challenges in COVID-19 vaccine distribution, while preserving patient privacy, and maximizing public engagement. In our recent work, we proposed a similar solution based on QR codes which enables it to function without an app \cite{bae2021mit}. However, our app based solution can facilitate zero knowledge verification, dynamic vaccine scheduling, continuous symptoms reporting, access to aggregate analytics based on population trends and more. Furthermore, our mobile app solution is aimed to work with limited access to the internet as well.



\section{Use cases and functional requirements of vaccine management platforms}

We identify six critical functions that must be performed by systems facing vaccine recipients in the continued distribution of COVID-19 vaccines. Each of these six critical functions addresses the important challenges in equitable vaccine distribution that were previously identified and described in \cite{bae2021mit}. 

\subsection{Vaccination eligibility verification}

The current phased vaccination schedules adopted by several national governments \cite{intlstrats} prioritize specific subsets of the population for vaccination, to maximize the societal benefit of initially limited vaccines. In the United States, healthcare workers and long-term care facility residents and workers will be the first to receive a vaccine, with the following phases including essential workers, seniors, and those with high-risk medical conditions  \cite{vaccinetrack}. 
Confirming that an individual is part of an eligible population will be important in preventing fraud and to ensure equitable vaccine distribution. Since some eligibility requirements are based on PII, this process must be done in a privacy preserving way to ensure that this data is not misused. 

\subsection{Vaccine scheduling and administration}
Coordinating the vaccination of large groups within small time windows will require efficient scheduling systems to optimize the number of individuals being vaccinated. These scheduling systems must be accessible and interoperable with an assortment of vaccination sites across a diverse range of populations and locations \cite{himss}. Although multiple such systems have been developed, none have been widely successful \cite{nh}.

\subsection{Second dose coordination and record linkage }
Most vaccine candidates likely to reach widespread distribution require a minimum one dose and maybe followed by a booster shot for full immunogenicity. For maximum public utility, it is imperative that individuals consistently receive these doses of a COVID-19 vaccine. 

Furthermore, because studies have not yet been performed to determine whether taking a first dose COVID-19 vaccine from one manufacturer can be effectively complemented by a second dose from another manufacturer, current guidelines recommend that individuals should receive both vaccine doses from a single manufacturer \cite{timing}. Effective systems must be implemented to facilitate this process and to monitor the relative efficacy of multi-manufacturer dose schedules if these guidelines are changed in the future. Record keeping and reminder systems must be developed to ease the logistical burden of coordinating adherence to these specific vaccine protocols \cite{reminders}. Also, the same would be applicable for coordination of booster shots and long term adverse effect tracking.

However, it is important to acknowledge that single-dose vaccines are in development, and the approval of such a vaccine would make this requirement completely obsolete \cite{jnj}.

\subsection{Vaccination verification}
As more of the population get vaccinated, methods of assessing and confirming vaccination status become increasingly important for the relaxation of certain public health measures and shutdowns \cite{vaccineneeds}. Therefore, before policies dependent on immunization status can be enacted, secure, fraud-resistant methods must be developed to confirm that an individual has been vaccinated.

\subsection{Safety efficacy monitoring}

The novelty of COVID-19 vaccine platforms as well as the limited Phase III clinical trial data on their long-term efficacy and side effects can be supplemented by long-term health data collection. This will be important in the development of vaccination policies, public health measures, and can further elucidate differences in side effect manifestation and efficacy in diverse populations \cite{book}. Furthermore, this data can be used for quality assurance processes and aid vaccine manufacturers in improving vaccine technologies. Several challenges exist in the secure, privacy-focused collection of this individual protected health information (PHI).   

\subsection{Trust and communication}
Every intersection between user engagement and any vaccine management system is an opportunity to build trust and provide transparent communication between all stakeholders involved in the vaccine distribution pipeline (coupon distributor, vaccinator, the guard at a venue, etc). Effective communication across these user-facing systems as well as thorough messaging surrounding their features and usage will be important in gaining public trust and increasing engagement \cite{euro}.
\section{Centralized Approaches with PII}

Centralized systems including the Vaccine Administration Management System (VAMS) and the V-Safe After Vaccination Health Checker platforms, are being developed by the government to address various required functions in vaccine administration \cite{VAMS}. The  Vaccine Adverse Event Reporting System (VAERS) is a long-standing centralized system for safety and efficacy monitoring as well. Some of these are developed in conjunction with private sector companies. The primary intended audience for these systems varies, and therefore there is also variance in user-experience.

\subsection{User Flow}

VAMS is a multifunctional platform for vaccine administration and monitoring meant to be used by healthcare providers, employers, vaccine clinic managers, and vaccine recipients. The system includes processes for vaccine eligibility prioritization, appointment scheduling, dose recording, second dose reminders, and vaccination status verification. The separate Vaccine Adverse Event Reporting System (VAERS) and V-safe After Vaccination Health Checker platforms enable the reporting of side-effects and symptoms that may be associated with vaccination \cite{VAERS}. 

\subsection{Challenges}
\subsubsection{Privacy and data security}
 VAMS requires the input of extensive protected health information (PHI) including questions concerning HIV status, cancer diagnosis, and other pre-existing conditions \cite{VAMSmanual}. The storage of this sensitive private data in a centralized manner opens the potential for the loss of personally identifiable information (PII) and PHI for large portions of the population if security breaches occur.

\subsubsection{Logistics and efficiency}
Due to requirements for the input of substantial amounts of information, usage of centralized government systems including VAMS, VAERS, and V-safe is an arguably time-intensive and complex process. This can create unnecessary friction in the vaccine distribution pipeline.

\subsubsection{Engagement and user-trust}
Existing barriers to user trust and current vaccine hesitancy trends may be exacerbated by centralized systems requiring the input of large amounts of PII. Officials have voiced concerns that the collection of this PII is unnecessary for the monitoring of health outcomes and efficacy, and may be used for other purposes such as the identification of undocumented immigrants \cite{cuomo}. This has the potential to discourage vaccination in minority populations that are already being disproportionately affected by COVID-19.

\subsubsection{Monitoring of health outcomes}
Though centralized systems promise to comprehensively monitor health outcomes by the collection of large volumes of information, there are challenges to the use of this information in improving continued vaccination efforts. Despite increased information collection, the potential decrease in user engagement can create a net deficit in reported vaccine side effects and efficacy when compared to lower-friction, privacy-focused solutions. Second, there has been no clear messaging on how collected data will be used and it is unclear whether vaccine manufacturers will be provided access to this information to continue to improve vaccine development \cite{natgeo}. 

\begin{figure}[ht!]
\begin{center}
\includegraphics[width=12cm]{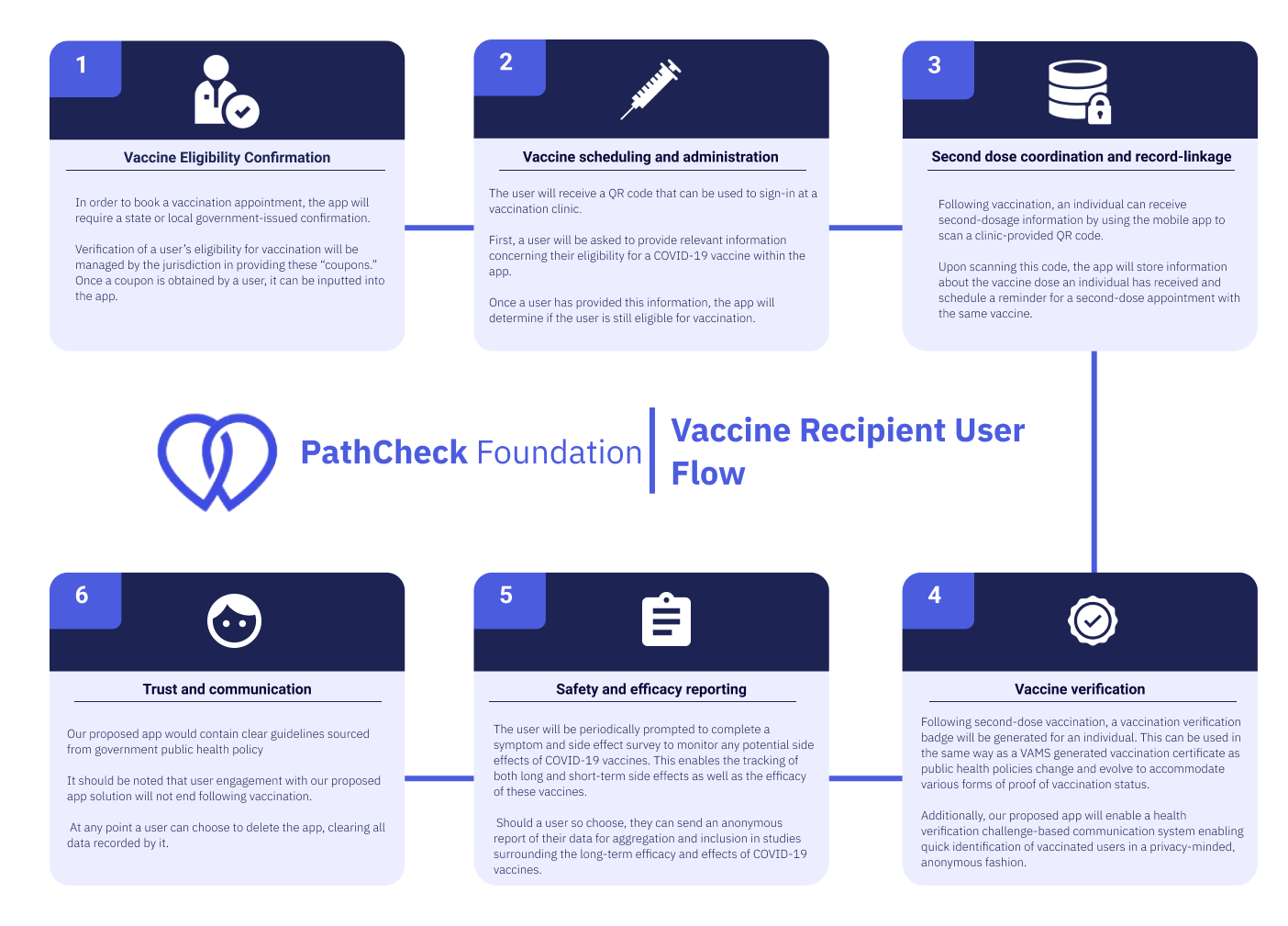}
\end{center}
\caption{Flowchart depicting the Vaccine Recipient User Flow.}
\label{flow}
\end{figure}
\section{An app-based approach}
\subsection{Vaccine Recipient User Flow}
Our proposed mobile app approach for vaccine coordination focuses primarily on user engagement and individual health outcomes. This app will be computationally lightweight while also fulfilling numerous functions including eligibility confirmation, vaccine schedule, side effect reporting, and vaccination verification. We also believe that it will be easily integrated into existing systems such as VAMS, ideally supplementing these approaches. 
 
Entry into our proposed app-based solution for vaccine distribution would be non-restrictive. Any individual has the opportunity to download the open-source app and can utilize features including the FAQ, symptom/vaccination dashboards, etc.

\subsubsection{Vaccine eligibility confirmation}
In order to book a vaccination appointment, the app will require a state or local government-issued confirmation. This might take the form of a physical vaccine “coupon” provided to individuals physically either through the mail, an employer, or at a government pickup site. Verification of a user’s eligibility for vaccination will be managed by the jurisdiction in providing these “coupons.” Once a coupon is obtained by a user, it can be inputted into the app. One-time passwords (OTPs) could be used to prevent access to digitally stored QR codes. At this point, our proposed app will not require the user to provide any personally identifiable information (PII). 

\subsubsection{Vaccine scheduling and administration}

After the verification of a valid vaccine coupon, the user has the ability to create a vaccination appointment. Because the user’s app has been verified by the input of a coupon, vaccination can be anonymized. Information about vaccination clinics and the vaccination methods that they support will be listed within the app (drive-through, walk-in, anonymous, etc.). Once a user has confirmed a vaccination appointment, they will receive a QR code (issued by CDC) that can be used to sign-in at a vaccination clinic. 

The process of receiving a vaccine can also proceed anonymously. First, a user will be asked to provide relevant information concerning their eligibility for a COVID-19 vaccine within the app. For instance, current COVID-19 vaccines have not yet been tested in children, pregnant women, and immunocompromised populations. Once a user has provided this information, the app will determine if the user is still eligible for vaccination. PHI and PII are not stored and are only used for eligibility confirmation. Once this information is confirmed, the app will generate a QR code for check-in at a vaccination clinic. It is at this point that our proposed vaccine app solution might interface with VAMS in order to supply basic vaccine information. No PII will be transferred, but information about the clinic, first or second dose status, and non-identifying information such as age-range, race, and sex may be communicated to VAMS for record-keeping purposes. A randomized user identifier will be transmitted to VAMS in the place of a user’s name. 

\subsubsection{Second dose coordination and record-linkage}

Following vaccination, an individual can receive second-dosage information by using the mobile app to scan a clinic-provided QR code. In some cases, this might be in the form of barcodes already being created by various vaccine manufacturers for each dose. Otherwise, these codes might be printed by vaccination clinics or provided on physical information sheets handed to recipients following vaccination. Upon scanning this code, the app will store information about the vaccine dose an individual has received and schedule a reminder for a second-dose appointment with the same vaccine. This reminder could be adjusted later based on the supply of doses. The vaccine app will also generate a unique vaccine stamp indicating that the user has received the first dose of a COVID-19 vaccine. Second-dose scheduling and vaccination follow the same workflow as above. 

\subsubsection{Vaccine verification}

Following second-dose vaccination, a vaccination verification badge will be generated for an individual. This can be used in the same way as a VAMS generated vaccination certificate as public health policies change and evolve to accommodate various forms of proof of vaccination status. 
Additionally, our proposed app will enable a health verification challenge-based communication system. This enables the quick identification of vaccinated users in a privacy-minded, anonymous fashion. If vaccination becomes a requirement for entry to venues or participation in events, this system can be used to quickly and privately identify vaccinated individuals.  

\subsubsection{Safety and efficacy reporting}
Following the receipt of a first-dose COVID-19 vaccine, a user will be periodically prompted to complete a symptom and side effect survey to monitor any potential side effects of COVID-19 vaccines. This enables the tracking of both long and short-term side effects as well as the efficacy of these vaccines. This data is not initially automatically uploaded from a user’s device. Instead, it is stored for an individual to personally use in symptom and side-effect tracking and for comparison against aggregated trends. Should a user so choose, they can send an anonymous report of their data for aggregation and inclusion in studies surrounding the long-term efficacy and effects of COVID-19 vaccines. We envision this anonymized data being provided to both vaccine manufacturers for quality control and vaccine technology improvement as well as government agencies for the continued development of public health policies. This anonymous data sharing is another point of potential contact between a decentralized app and systems such as VAERS or V-safe. At this stage, a user is still not required to input PII but can choose to do so if they desire a response from a government agency regarding any side effects or symptom concerns. Users can also receive notifications regarding safety and efficacy via push notifications rather than via personal contact information. 

\subsubsection{Trust and communication }
Our proposed app would contain clear guidelines sourced from government public health policy. Further, the open-source, open standards nature of the app allows careful examination by the public to build user trust. It should be noted that user engagement with our proposed app solution will not end following vaccination. Continuing users will continue to be periodically prompted to input updates to their health status and any side effects that they may experience. This longitudinal platform can help supplement efforts to study and monitor the long-term efficacy and side effects of COVID-19 vaccines. At any point a user can choose to delete the app, clearing all data recorded by it. 

\subsection{Challenges}
\subsubsection{Privacy and data security}
Our proposed decentralized solution is intended to be exceedingly privacy-focused with little opportunity for data breaches or spoofing. The primary potential vulnerabilities exist only if a user should misplace their smartphone or with the possibility of targeted cyber attacks on the phone itself. However, because there is no centralized repository for personal information, these attacks cannot result in the loss of multiple individuals’ records. 
\subsubsection{Logistics and efficiency}
For an individual user, an app-based solution would be designed for convenience and user experience. We have outlined several points at which such an app might be interoperable with centralized systems such as VAMS and VAERS, but there still may be challenges in developing cooperative relationships enabling this interoperability. Nonetheless, our proposed app would be a highly efficient single source of all vaccine-related functionality for an individual who needs to be vaccinated.  

\subsubsection{Engagement and user-trust}
A primary difficulty in adoption of a vaccine management app is ineffective public messaging that does not detail the privacy-minded characteristics [14]. User engagement and trust will hinge largely on effective communication and education surrounding the privacy and ease of use of app based solutions and the many privacy-preserving features they enable in contrast to existing centralized systems such as VAMS. The open-source nature of our approach enables a comprehensive description of features and security measures with complete transparency, which may be complemented with effective educational messaging to increase understanding of the relative benefits of app-based solutions. 

\subsubsection{Monitoring of health outcomes}
While this decentralized app-based solution may solve several challenges faced by centralized systems with respect to privacy, convenience, and user engagement, it may still be difficult to properly incentivize users to report symptoms. Furthermore, should the app use anonymized data in aggregate for statistical purposes, there may be hesitancy and privacy concerns despite the lack of collected PII. 

\section{Conclusion}
The high volume PII requirements of VAMS facilitate country-level insights into vaccine distribution at a granular level stratified by a variety of individual identifiers. However, several challenges also exist in the widespread adoption of these systems due to concerns over data privacy and barriers to ease of use of these systems. Here we propose an app-based solution for vaccine coordination that can supplement centralized approaches and is user-focused, privacy-preserving, efficient, and easily scalable. 

Key characteristics of our proposed vaccine monitoring app supplement include the opportunity for anonymity and privacy in most steps of the vaccination process. Furthermore, this app consolidates each of the six most critical functions in COVID-19 vaccination into a single platform, streamlining the user experience and ideally increasing engagement. Our app also provides the opportunity for large scale, participatory data collection efforts to monitor side-effects and efficacy of COVID-19 vaccines without compromising privacy, ideally providing a source of de-identified health information for government policy development and vaccine manufacturer insights. 

We also propose our SafePaths card-based approach to extend privacy-preserving vaccine management systems to low resource areas and those without access to mobile devices. 

Crucially, our privacy-preserving app and SafePaths card systems would be interoperable with other vaccine coordination monitoring and administration platforms such as VAMS and VAERS, reducing the difficulty of integrating this new approach. We believe that this app-based approach can incentivize vaccination for individuals valuing personal privacy including undocumented immigrant populations and those with pre-existing health conditions.

\subsubsection*{Acknowledgments}
We are grateful to Riyanka Roy Choudhury, CodeX Fellow, Stanford University, Adam Berrey, CEO of PathCheck Foundation, Dr. Brooke Struck, Research Director at The Decision Lab, Canada, Vinay Gidwaney, Entrepreneur and Advisor, PathCheck Foundation, and Paola Heudebert, co-founder of Blockchain for Human Rights, Alison Tinker, Saswati Soumya, Sunny Manduva, Bhavya Pandey, and Aarathi Prasad for their assistance in discussions, support and guidance in writing of this paper.

\bibliography{iclr2021_conference}
\bibliographystyle{iclr2021_conference}

\section*{Frequently Asked Questions}

\subsection*{What are the benefits of this app for various stakeholders?}
\textit{For public health agencies: }
\begin{enumerate}
    \item \textbf{Interoperability:} Integrates with existing systems with processes for easy information sharing.
    \item \textbf{Coordination and Efficiency:} Reduced dependency on digital IT systems working seamlessly with one another (hospital EMR systems, clinic management systems, centralized monitoring systems).
    \item \textbf{Feedback and monitoring:} App-based reporting of symptoms and efficacy directly from vaccinated individuals can be reported in nearly real-time. 
    \item \textbf{Communication:} Possibility for targeted, contextual messaging and information sharing. 
    \item \textbf{Data aggregation and dashboard visualization:} Granular data anonymously reported from users can be used to monitor population-level trends. 
\end{enumerate}

\textit{For users:}
\begin{enumerate}
    \item \textbf{Privacy:} End-to-end system for vaccination without requiring the sharing of PII. 
    \item \textbf{Equity:} Allows anonymous access to all marginalized and vulnerable populations. The VaxSafe card system extends this to those without mobile app access.
    \item \textbf{Efficiency:} Single platform for all functions of COVID-19 vaccination. Reduced requirement for redundant and intrusive data input.  
\end{enumerate}

\textit{For pharmaceutical companies and vaccine makers: }
\begin{enumerate}
    \item \textbf{Quality assurance:} Clear pipeline for side-effect data sharing that can supplement adverse event reports from VAERS.
    \item \textbf{Monitoring and Reporting:} Opportunity for collaboration to build systems reporting the most relevant efficacy and side-effect information to monitor long-term side effects and efficacy. 
    \item \textbf{Messaging:} Opportunity for ‘recall’ of vaccine lots with reduced efficacy and ability to alert affected users.
\end{enumerate}

\subsection*{How will we encourage users to download the app?}
First, we believe that the benefits of a privacy-focused app will attract users to prioritize data-security in the COVID-19 vaccination process. We also believe that the efficiency and ease of use of our proposed vaccination framework will appeal to many individuals. Additionally, we plan to tap into an already large existing user-base of EN contact tracing app users. Finally, we will build on existing partnerships with medical centers and state governments. 

\subsection*{Is this vaccine app a replacement for existing CDC or state systems?}
No. Systems such as VAMS/VAERS/V-Safe/IIS/IZ will still be important in widespread COVID-19 vaccination efforts. Our proposed app would provide an alternative method for vaccination that is privacy-sparing, efficient, and equitable while serving as a supplemental source of vaccine monitoring information.  

\subsection*{How will this interface with VAMS/VAERS/V-Safe/ IIS/IZ? What changes are required?}
Vaccine diary, second-dose, and health status alert, and informational features of our proposed app would be independent of existing systems. The input of vaccination information upon administration of a vaccine and side effect/efficacy reports are two areas with potential for integration with government systems. 

To verify and record vaccine administration, vaccination clinics or governments would need to provide signed QR codes that can be printed/copied by pharmacies or by users. This QR code would have information regarding the lot, manufacturer, and dosing of a vaccine which can then be verified by others with the appropriate digital key. 

For interoperability of symptom/side effect reports, state or federal systems will need to allow the pseudorandom identifier associated with a user to be used for data identification purposes rather than PII such as name, address, etc. This is already part of the PPRL (privacy-preserving record linkage) protocol for VAMS and IIS.

\subsection*{If you don't have PII, how can a doctor or healthcare provider get in touch with the user?}
Doctors and public health officials can contact users regarding pertinent information about their specific vaccine lot and other important details via app-mediated push notifications and contextual alerts. This is similar to ‘recalls’ in auto-parts, food safety, toys, etc. 

\subsection*{What difference will it make? Wouldn’t everyone be vaccinated anyway?}
Significant chunks of the population still exhibit vaccine hesitancy and many may be unwilling to receive a COVID-19 vaccine. This app aims to address potential barriers to vaccination by protecting data privacy, creating a convenient, streamlined user experience, and providing multiple vaccine-related functionalities in one platform. 

\subsection*{Is this app primarily a vaccine passport or verifiable credentials?}
This app does support vaccine verification while also including modules surrounding eligibility confirmation, dose scheduling and reminders, health assessments, and symptom reporting, and providing users with push-notifications and contextual alerts. 

\subsection*{How will you reach marginalized and low-resource communities?}
We have also proposed state-produced physical vaccine cards that can be used for many of the functions of our app solution. This enables a privacy-focused solution for vaccination. Please see a thorough explanation in section 6. 

\subsection*{Why should users trust the app?}
Our app is developed using open-source code and open standards. 

\subsection*{Why do centralized systems including VAMS and VAERS require so much PII and HPI?}

PII including name, date of birth, and contact information is primarily used for user identification, contact, and record-keeping. Health information is stored to determine eligibility for vaccination based upon exclusion criteria and to track the interactions between various medical conditions and vaccination. Other personal information might be used for aggregate analysis and statistical purposes regarding equitable distribution among diverse populations. Our app-based approach addresses each of these functions without the use of PII. 

Identification of an individual for record-keeping is performed using a pseudorandom identification number rather than name or date of birth. Previous health information can be inputted into the app for exclusion determination where it is not stored. As soon as the app determines eligibility for vaccination information shared in these questions will be deleted. Symptom and adverse event reporting can be performed either anonymously or with personal information that might lend insight into the vaccine and medical condition interactions. All demographic information can be anonymized and aggregated for reporting. 

\subsection*{What if the user does not have a smartphone?}
We expect users seeking a privacy-oriented approach to vaccination to use a physical card containing a digitally-signed QR code from the government. 

\subsection*{What is PathCheck and what role can it play?}
PathCheck is a nonprofit organization originating in Dr. Ramesh Raskar’s lab at MIT. We are the world’s largest open-source, open standards non-profit organization for COVID-19 and research a broad array of problems stemming from the pandemic. PathCheck was the first organization to launch an EN app for contact tracing in COVID-19, successfully partnering with 6 US states and territories. 

\subsection*{What is MIT SafePaths? What is its role?}
MIT SafePaths is a set of standards protocols and algorithms. These are included as part of the open-source tools developed by MIT to help combat the COVID-19 pandemic.

\end{document}